\documentclass[prb,twocolumn,showpacs,floatfix,amsmath,amssymb,superscriptaddress]{revtex4}
\usepackage[dvips]{graphicx}
\usepackage{latexsym}
\usepackage{graphicx}
\usepackage{times}
\usepackage{amsmath}
\usepackage{dcolumn}
\usepackage{latexsym,amsmath,amssymb,bm,euscript}
\begin{document}

\title{Graphene integer quantum Hall effect in the ferromagnetic 
 and paramagnetic regimes}

\author{Jason Alicea}
\affiliation{Physics Department, University of California, 
Santa Barbara, CA 93106}
\author{Matthew P. A. Fisher}
\affiliation{Kavli Institute for Theoretical Physics, 
University of California, Santa Barbara, CA 93106}

\date{\today}

\begin{abstract}

Starting from the graphene lattice tight-binding Hamiltonian with an
on-site $U$ and long-range Coulomb repulsion, we derive an interacting 
continuum Dirac theory governing the low-energy behavior of graphene
in an applied magnetic field.  Initially, we consider a
clean graphene system within this effective theory and
explore integer quantum Hall ferromagnetism
stabilized by exchange from the long-range Coulomb repulsion.  
We study in detail the ground state and excitations at $\nu =
0$ and $\nu = \pm 1$, taking into account small symmetry-breaking
terms that arise from the lattice-scale interactions, and also explore
the ground states selected at $\nu = \pm 3$, $\pm 4$, and $\pm 5$.  
We argue that the ferromagnetic regime may not yet be realized in 
current experimental
samples, which at the above filling factors perhaps remain
paramagnetic due to strong disorder.  In an attempt to
access the latter regime where the role of exchange is strongly
suppressed by disorder, we apply Hartree theory to study the effects of
interactions.  Here, we find that Zeeman splitting together with 
symmetry-breaking interactions can in principle produce integer 
quantum Hall states in a paramagnetic system at $\nu = 0$, $\pm 1$ 
and $\pm 4$, but not at $\nu = \pm 3$ or $\pm 5$, consistent 
with recent experiments in high magnetic fields.  We make 
predictions for the activation energies in these 
quantum Hall states which will be useful for determining their true origin.

\end{abstract}

\maketitle


\section{Introduction}
\begin{figure} 
  \begin{center} 
    {\resizebox{6cm}{!}{\includegraphics{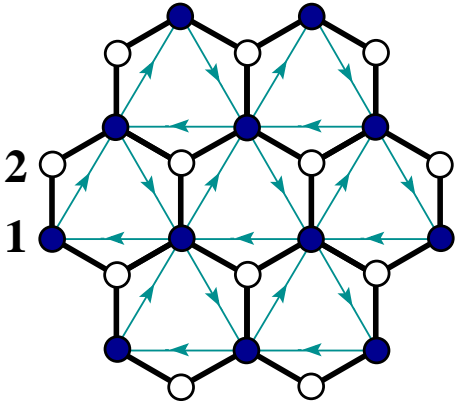}}} 
  \end{center} 
  \caption{Schematic lattice-scale order in the $\nu = -1$ integer quantum
  Hall state.  ``1'' and ``2'' label the two
  sublattices of the honeycomb.  Electrons in the highest-occupied
  Landau level reside only on sublattice 1 (for instance), and undergo
  circulating currents around second-neighbor plaquettes
  oriented along the arrows.  } 
  \label{honeycomb} 
\end{figure}

Recent experimental advances have made possible 
the isolation of high quality two-dimensional graphene
sheets,\cite{GrapheneFabrication} thus opening up a new arena
for exploring quantum Hall physics.  As illustrated
in Fig.\ \ref{honeycomb}, graphene is a collection of carbon atoms arranged
on a honeycomb lattice.  At half-filling and in the absence
of a magnetic field, graphene is semi-metallic with a Fermi surface
consisting of two distinct nodes residing at the Brillouin zone
corners.  
Low-energy excitations about the two nodes are well
characterized by two ``flavors'' of Dirac fermions whose energies obey 
a linear dispersion relation, $E({\bf k}) = \pm\hbar v|{\bf k}|$,
where for graphene $v \approx 10^6$m/s.\cite{IQHE1,IQHE2}   
Applying a
magnetic field rearranges the spectrum into Landau
levels, each of which is approximately four-fold degenerate near the
Fermi level owing to the
presence of flavor and spin.  The underlying Dirac structure gives
rise to an unconventional quantum Hall effect which has been observed
experimentally \cite{IQHE1,IQHE2,IQHE3,HighFieldExpt} and 
explored theoretically from numerous perspectives.\cite{Zheng,
  Gusynin, Peres1, Peres2, McCann, Sheng, Brey, Levitov, Lukose, 
  GrapheneMacDonald, FertigBrey}

Initial experiments on the graphene quantum Hall effect reported
quantized Hall plateaus at $\sigma_{xy} = \nu e^2/h =
4(n+1/2)e^2/h$, where $n$ is
an integer.\cite{IQHE1,IQHE2}  The appearance of these integer
quantum Hall states is quite natural from a non-interacting, 
single-particle perspective, given the approximate four-fold Landau
level degeneracy.  There are, however, at least two mechanisms by which
additional integer quantum Hall states in graphene may be induced.  First, such
states can in principle be stabilized by explicit symmetry-breaking 
fields and/or interactions, the simplest mechanism being Zeeman
splitting.  In this situation the system would be appropriately
characterized as a ``quantum Hall paramagnet''.  Second, provided disorder 
is sufficiently weak 
the poorly screened Coulomb interactions in graphene can  
dramatically modify the single-particle picture,
in analogy with the well studied ``quantum Hall ferromagnetism'' in
GaAs heterostructures.\cite{GirvinReview}  In the latter system, for
instance, exchange interactions strongly stabilize the integer quantum
Hall state at $\nu = 1$, even in the limit of vanishing Zeeman
coupling --- that is, the ground state is a
ferromagnet.  Generally speaking, this mechanism is actually far 
more common than the
first since the Coulomb energy scale is typically much
larger than the energy scale associated with symmetry-breaking terms
such as the Zeeman energy.  Additional exchange-driven integer 
quantum Hall states can similarly be expected to appear in a 
sufficiently clean graphene system.  
In fact, quantum Hall ferromagnetism in graphene promises to be even
richer than in GaAs due to the additional flavor degree of
freedom, which as we will discuss offers the interesting possibility
of lattice-scale order coexisting with the integer quantum Hall effect.
The problem is thus more analogous to quantum Hall
bilayer and Si MOSFET physics,\cite{BilayerReview} where layer 
and valley indices play a role similar to flavor in graphene.  

Interestingly, more recent experiments utilizing higher magnetic
fields have in fact resolved additional quantized Hall plateaus at
filling factors $\nu = 0, \pm 1$, and $\pm 4$.\cite{HighFieldExpt}  
The activation energy measured at $\nu = \pm 4$ 
was found, somewhat surprisingly, to be dominated by the 
single-particle Zeeman energy,
suggesting that at these filling factors the system resides in the
paramagnetic regime due to strong disorder.  
The origin of the quantum Hall states at $\nu = 0$ and
$\nu = \pm 1$ is unclear at present, and presents an interesting
puzzle.  In a very recent paper Nomura
and MacDonald suggest that these states are due to the onset of quantum Hall
ferromagnetism.\cite{GrapheneMacDonald}  Given the apparent spin-splitting 
origin of the states at $\nu = \pm 4$, however, it is perhaps worth
exploring an alternative scenario where the quantum Hall states at
$\nu = 0$ and $\nu = \pm 1$ are similarly due to explicit symmetry
breaking, rather than Coulomb exchange.  

The purpose of the present paper is twofold.  In the first part, we
carry out a detailed exploration of quantum Hall ferromagnetism,
focusing on a clean graphene sample.  Following this analysis, we 
consider a dirty system
and ask whether symmetry-breaking terms can stabilize integer
quantum Hall states in the paramagnetic regime.  
To this end, we start with the lattice
Hamiltonian including on-site $U$ and long-range Coulomb interactions,
and derive a continuum interacting Dirac formulation suitable for
studying these phenomena.  In
addition to the usual long-range Coulomb repulsion, we retain
shorter-range terms that arise from the microscopic interactions and
explicitly break the flavor degeneracy exhibited in a non-interacting
theory.  Such symmetry-breaking terms have not been taken into account
previously, and are important for analyzing the physics both in the
ferromagnetic and paramagnetic limits.    
In the ferromagnetic regime, we analyze the ground state and
excitations within this theory at integer filling factors in the
lowest two Landau levels.  At filling factors $\nu = \pm 1$,
we show that interactions favor ``easy-axis'' flavor polarization, 
leading to a ground state exhibiting charge density wave order on the 
lattice scale.  The $\nu = 0$ ground state is found to depend on the
strength of interactions relative to the Zeeman coupling, and will
either be a uniform spin-polarized state or a spin-singlet
exhibiting the same
lattice-scale structure as at $\nu = \pm 1$.  We calculate the
spin-wave and particle-hole excitation energies at these filling
factors, as well as identify the relevant skyrmions which are
expected to set the transport gap.  Interestingly,
interactions favor ``easy-plane'' flavor polarization at $\nu = \pm
3$ and $\pm 5$, leading to a ground state that spontaneously breaks
U(1) flavor symmetry.  Thus, a finite-temperature 
Kosterlitz-Thouless transition can be expected at these filling
factors.  Finally, a uniform, spin-polarized ground state 
is expected at $\nu = \pm 4$.  We 
establish that skyrmions continue to provide the minimum-energy charge 
excitations at $\nu = \pm 3, \pm 4$, and $\pm 5$ as well, at least in 
the absence of anisotropy terms.

To explore the paramagnetic regime in a dirty system, we use the 
expectation that the effects of exchange should be strongly suppressed by 
disorder and incorporate interactions within Hartree theory.  
Here, we argue that in principle interactions can give
rise to quantum Hall states at $\nu = 0, \pm 1$, and $\pm 4$, but
not at $\nu = \pm 3$ or $\pm 5$, which is consistent with the recent
high-field experiments.  We provide estimates for the activation
energies in these additional quantum Hall states, which will be useful
for determining experimentally whether the $\nu = 0$ and $\pm 1$ states 
originate from explicit symmetry breaking or quantum Hall ferromagnetism.

\section{Continuum Interacting Theory}
We start with the zero-field lattice Hamiltonian for graphene 
written as 
\begin{equation}
  H = H_t + H_U + H_{\rm Coul}, 
\end{equation}
where $H_t$ describes hopping
of electrons across nearest-neighbor honeycomb sites,
\begin{equation}
  H_t = -t\sum_{\langle {\bf x x'}\rangle}\sum_{\alpha =
  \uparrow,\downarrow} [c^\dagger_{\alpha {\bf x}}
  c_{\alpha {\bf x'}} + {\rm H.c.}],
\end{equation}
and $H_U$ and $H_{\rm Coul}$ contain the on-site repulsion and
long-range Coulomb interaction, respectively:
\begin{eqnarray}
  H_U &=& U \sum_{\bf x}\bigg{[}\frac{1}{4}(n_{\bf x})^2
  -\frac{1}{3}{\bf S}({\bf x})^2\bigg{]}
  \\
  H_{\rm Coul} &=& \frac{1}{2}\sum_{\bf x \neq x'} V({\bf x -
  x'})n_{\bf x}n_{\bf x'}.
\end{eqnarray}
Here $n_{\bf x} = n_{\uparrow {\bf x}} + n_{\downarrow {\bf x}}$ is the
electron number operator and ${\bf S}({\bf x}) = \frac{1}{2}
c^\dagger_{\alpha {\bf x}}{\bm \sigma}_{\alpha\beta} c_{\beta {\bf
x}}$ is the usual spin operator with ${\bm \sigma}$ a vector of Pauli
matrices.  The Coulomb potential is
$V({\bf x}) = \frac{e^2}{4\pi \epsilon}\frac{1}{|{\bf x}|}$, with
$\epsilon$ an
appropriately chosen dielectric constant (see below).  We will
introduce the magnetic field upon entering the continuum, which is
justified since for experimental field ranges
the magnetic length $\ell_B = \sqrt{\hbar/(eB)}$ is much larger than
the lattice spacing $a_0$.  

As mentioned above, with one electron per site, the honeycomb
band structure exhibits two gapless
Dirac points at the Fermi energy, occurring at wave vectors $\pm {\bf Q}
= \pm (4\pi/3,0)$.  Focusing on the linearly dispersing excitations
near the two nodes, the Fourier-transformed lattice fermion 
operators may be conveniently expanded in terms of two flavors
of continuum Dirac fermion fields (denoted $R$ and $L$) as follows,
\begin{eqnarray}
  c_{\alpha {\bf q+Q} a} &\sim& \gamma \psi_{\alpha R a}({\bf q})
  \label{DiracFields1}
  \\
  c_{\alpha {\bf q-Q} a} &\sim& \gamma i \eta^y_{ab} \psi_{\alpha L b}({\bf
  q}).
  \label{DiracFields2}
\end{eqnarray}
Here and below we reserve indices $\alpha,\beta$ for spin, $A,
B$ for flavor, and $a,b$ for the honeycomb sublattice.
Moreover, $\sigma^j_{\alpha \beta}$, $\tau^j_{AB}$, and $\eta^j_{ab}$ 
denote Pauli
matrices that contract with the spin, flavor, and sublattice indices,
respectively.  (Note that according to Eqs.\ (\ref{DiracFields1}) and
(\ref{DiracFields2}) 
$\psi_{R1}$ and $\psi_{L2}$ correspond to sublattice 1, while
$\psi_{R2}$ and $\psi_{L1}$ correspond to sublattice 2.)
We will use the convention that suppressed indices on the fields 
are implicitly summed
(\emph{i.e.}, $\psi^\dagger \psi \equiv \sum_{\alpha A
  a}\psi^\dagger_{\alpha A a } \psi_{\alpha A a}$).  
The normalization on the right side of Eqs.\ (\ref{DiracFields1}) and
(\ref{DiracFields2}) is chosen to be $\gamma =
\sqrt{2}/(3^{1/4}L)$ so that the continuum fields satisfy the
canonical anticommutation relations $\{\psi_{\alpha A a}({\bf
  q}),\psi^\dagger_{\beta B b}({\bf q'})\} = \delta_{\alpha \beta} \delta_{AB}
\delta_{ab} (2\pi)^2 \delta^2({\bf q-q'})$.
Real-space Dirac fields are defined according to
\begin{eqnarray}
  \psi_{\alpha A a}({\bf q}) &=& \int d^2{\bf x} 
  e^{-i {\bf q \cdot x}} \psi_{\alpha A a}({\bf x})
\end{eqnarray}
and satisfy 
$\{\psi_{\alpha A a}({\bf
  x}),\psi^\dagger_{\beta B b}({\bf x'})\} = \delta_{\alpha \beta} \delta_{AB}
\delta_{ab}\delta^2({\bf x-x'})$.  Note that consistency requires 
$\psi({\bf x})$ to be a slowly varying field, or equivalently that
$\psi({\bf q})$ is nonzero only within some momentum-space cutoff
$\Lambda$.  The lattice electron number operator $n_{{\bf x}a}$ for
sublattice $a$ can now be expressed in terms of real-space Dirac
fields as follows,
\begin{eqnarray}
  n_{{\bf x} 1} &=& \frac{\sqrt{3}a_0^2}{2}[\rho_{R 1} +
  \rho_{L 2} 
  + e^{2i{\bf Q\cdot x}} J_+^\dagger + e^{-2i{\bf Q\cdot x}} J_+]
  \label{density1}
  \\
  n_{{\bf x} 2} &=& \frac{\sqrt{3}a_0^2}{2}[\rho_{R 2} +
  \rho_{L 1} 
  - e^{2i{\bf Q\cdot x}} J_-^\dagger - e^{-2i{\bf Q\cdot x}} J_-],
  \label{density2}
\end{eqnarray}
where $\rho_{A a} = \sum_{\alpha = \uparrow,\downarrow}
\psi^\dagger_{\alpha A a} \psi_{\alpha A a}$ contain uniform pieces of
the density and $J_+ =
\psi^\dagger_{R1}\psi_{L2}$ and $J_- = \psi^\dagger_{R2}\psi_{L1}$
contain oscillatory components of the density. 
The spin operators on each sublattice take on similar forms, with
$\frac{1}{2}{\bm \sigma}$ inserted between each 
$\psi^\dagger$ and $\psi$ above.  It follows that in the
continuum, the uniform part of the total density is 
$\rho_{\rm tot} = \psi^\dagger \psi$, while the uniform spin density is
${\bm S}_{\rm tot} = \frac{1}{2} \psi^\dagger {\bm \sigma} \psi$.

Using the continuum expansion above and turning on the external magnetic
field, we arrive at the desired continuum Hamiltonian, which we write as
\begin{equation}
  {\mathcal H} = {\mathcal H}_0 + {\mathcal H}_1.  
\end{equation}
Here ${\mathcal H}_0$ denotes part of the Hamiltonian that is 
invariant under global $SU(4) = SU(2)_{\rm spin} \times
SU(2)_{\rm flavor}$ rotations,
\begin{eqnarray}
  {\mathcal H}_0 &=& -i \hbar v \int d^2{\bf x}
  \psi^\dagger [\eta^x D_x+\eta^y D_y]\psi 
  \nonumber \\
  &+& \frac{1}{2} \int d^2{\bf x}d^2{\bf x'} 
  \rho_{\rm tot}({\bf x}) V({\bf x-x'}) \rho_{\rm tot}({\bf x'}),
  \label{H0}
\end{eqnarray}
with $D_j = \partial_j - i (e/\hbar) A_j$.  
${\mathcal H}_1$ encodes the remaining anisotropy terms which 
break this $SU(4)$ symmetry down to 
$U(1)_{\rm spin} \times[U(1) \times Z_2]_{\rm flavor}$, 
\begin{eqnarray}
  {\mathcal H}_{1} &=& \int d^2 {\bf x} \big{\{}- g \mu_B {\bf
  B}\cdot {\bf S}_{\rm tot}
  + \frac{1}{4}u_0 [\rho_{\rm tot}^2 
  +\rho_{\rm stag}^2
  \nonumber \\
  &-& \frac{8}{3}({\bf S}_{R1}^2+{\bf S}_{L2}^2 + 6 {\bf
  S}_{R1}\cdot {\bf S}_{L2} + (1\leftrightarrow 2)) ]
  \nonumber \\
  &-& \sum_{\bf r} v_1({\bf r}) 
  \rho_{\rm stag}({\bf x+r})\rho_{\rm stag}({\bf x}) 
  \nonumber \\
  &-& u_2 [J_+^\dagger J_+ + J_-^\dagger J_-^\dagger]\big{\}},
  \label{H1}
\end{eqnarray}
where 
$\rho_{\rm stag} = \psi^\dagger \tau^z \eta^z \psi$ represents the
staggered electron density between sublattices 1 and 2 of the
honeycomb and ${\bf S}_{Aa} = \frac{1}{2}
\psi^\dagger_{Aa} {\bm \sigma} \psi_{Aa}$.  
The sum in Eq.\ (\ref{H1}) is over 
triangular lattice vectors ${\bf r} = m {\bf e}_+ +
n {\bf e}_-$, where $m, n$ are integers and ${\bf e}_{\pm} = (a_0/2)(\pm
1,\sqrt{3})$.
The first two lines of Eq.\ (\ref{H1}) contain the Zeeman coupling and
the continuum form of the on-site $U$, with 
\begin{equation}
  u_0 = \sqrt{3} a_0^2 U/4.  
  \label{u0}
\end{equation}
For concreteness we take the magnetic field ${\bf B}$ perpendicular to the
graphene plane along the $+{\bf \hat z}$ direction throughout.
Apart from the long-range density-density repulsion exhibited in
${\mathcal H}_0$, the lattice Coulomb interaction also gives rise to 
the shorter-range terms in
the last two lines of Eq.\ (\ref{H1}), reflecting
lattice-scale physics.  The $v_1$ term represents an 
inter-sublattice repulsion that 
reflects the smaller Coulomb energy cost
for electrons residing on the same sublattice versus opposite 
sublattices.
The $u_2$ term represents the intra-sublattice repulsion between 
oscillating components of the density in Eqs.\ 
(\ref{density1}) and (\ref{density2}).
The coupling constants for these terms are given by
\begin{eqnarray}
  v_{1}({\bf r}) &=& \frac{\sqrt{3}a_0^2}{8} [ V({\bf r}
  + 1/\sqrt{3}{\bf \hat y}) - (1-\delta_{{\bf r},{\bf 0}})V({\bf r})]
  \label{v1}
  \\
  u_2 &=& -\sum_{\bf r \neq 0}\frac{\sqrt{3}a_0^2}{2} V({\bf
  r})\cos{\bf Q \cdot r}\approx \frac{4}{3}
  a_0^2 \bigg{(}\frac{e^2}{4\pi\epsilon a_0}\bigg{)}.
  \label{u2}
\end{eqnarray}
For later use we note that 
\begin{eqnarray}
  u_1 &\equiv& \sum_{\bf r} v_1({\bf r}) \approx \frac{\sqrt{3}}{4} u_2.
  \label{u1}
\end{eqnarray}

Some comments are worthwhile here.  
The above continuum formulation is expected to remain appropriate 
out to a cutoff $\Lambda \sim \pi/(4 a_0)$ or so from 
the Dirac nodes, which constitutes an appreciable fraction of the
Brillouin zone.  Elimination of modes outside of this radius will 
renormalize the parameters in the theory somewhat, though for simplicity we
have retained the bare values for the coupling constants.  
Furthermore, other symmetry-allowed interactions will in principle be 
generated as well; however, we shall assume these are subdominant.  
We also remark that in Eq.\ (\ref{H1}) we have only retained 
terms arising from the lattice Coulomb repulsion that are 
nonvanishing upon assuming a local
form in the continuum (namely, $v_1$ and $u_2$); other terms are
expected to essentially average out and will thus be unimportant.  
Given that $\psi$ is a slowly varying field, it is tempting to
approximate $v_1$ as a local interaction (as we have done for $u_2$).  
For most purposes this is in fact adequate, and we shall often make
this assumption to simplify our analysis.  However, as we will 
discuss below, despite appearances assuming 
a purely local form of $v_1$ leaves the SU(2) flavor symmetry 
unbroken in the ferromagnetic ground state at filling factors 
$\nu = \pm 1$, which is why we have retained the finite range of this
interaction.

We now turn to screening in the Dirac theory.  Since the $v_1$ and
$v_2$ terms are effective only on short length scales, these interactions are
essentially unmodified by screening, and we thus assume that the 
bare dielectric constant for graphene appears in 
Eqs.\ (\ref{v1}) and (\ref{u2}).  With air on one side of
the graphene plane and SiO$_2$ on the other, the unscreened dielectric 
constant is estimated to be $\epsilon \approx [(\epsilon_+^{-1} +
\epsilon_-^{-1})/2]^{-1} \approx 1.6\epsilon_0$, where 
$\epsilon_+ \approx \epsilon_0$
and $\epsilon_- \approx 4 \epsilon_0$ correspond to air 
and SiO$_2$, respectively.  The long-range part of
the Coulomb interaction in ${\mathcal H}_0$ 
is, however, expected to be (weakly) screened.
To estimate the screening, we first
note that the presence of the applied magnetic field is expected to become
important on length scales longer than roughly the magnetic length,
which is much larger than the inverse cutoff $\Lambda^{-1}$.
Thus the screened dielectric constant for the zero-field case should
not be dramatically modified by the presence of the field.  We
therefore use the zero-field screened dielectric constant for the
long-range Coulomb repulsion in Eq.\ (\ref{H0}), 
obtained within the random phase
approximation \cite{Gonzalez} which yields 
$\epsilon_{\rm RPA} \approx \epsilon + e^2/(8\hbar v)\approx 5
\epsilon_0$.

\section{Overview of Dirac Landau levels}
\label{LLoverview}

Before analyzing the full interacting theory, it will be useful to
briefly recall the Landau level spectrum for the non-interacting
case.\cite{Zheng, Haldane}  Throughout the paper we
implicitly work in the symmetric gauge.  As a consequence of the 
Dirac structure, the Landau level energies in the non-interacting
theory are given by
\begin{equation}
  E_{\uparrow/\downarrow ,n} = \mp \frac{1}{2} g \mu_B B + {\rm sign}(n) 
  \sqrt{2 e \hbar v^2 B |n|},
  \label{LLenergies}
\end{equation}
where $n$ is an integer.  Zeeman coupling weakly spin splits the
Landau levels, leaving a two-fold flavor degeneracy in the absence of
interactions.  
The two-component wave function for spin $\alpha$ and flavor $A$ 
in the $n^{th}$ Landau level can be written
\begin{equation}
  \Phi^{n}_{\alpha A} = 
  \begin{bmatrix}
    \phi^{n} \\ \chi^n
  \end{bmatrix}.
  \label{wavefunction}
\end{equation}
For $n \neq 0$, the elements $\phi^n$ and $\chi^n$ 
are simply wave functions for Schrodinger equation Landau levels with
indices $|n|-1$ and $|n|$, respectively.  In the symmetric gauge, the
elements are related by $\chi^n = |n|^{-1/2} {\rm sign}(n)
a^\dagger \phi^n$, where $a^\dagger$ is the usual raising
operator for Schrodinger equation Landau levels.
One can show that the total probability weight is divided equally
between the upper and lower elements of the wave function;
in other words, an electron with wave function $\Phi^{n \neq
  0}_{\alpha A}$ is equally likely to be found on either honeycomb
sublattice.  In contrast, the $n = 0$
Landau level wave functions have $\phi^{n = 0} = 0$, 
while $\chi^{n = 0}$ is a Schrodinger equation lowest Landau 
level wave function.  
By examining Eqs.\ (\ref{DiracFields1}) and (\ref{DiracFields2}) we see that 
$\Phi^{n = 0}_{\alpha R}$ has weight only on sublattice
2 of the honeycomb, while $\Phi^{n = 0}_{\alpha L}$ has weight only on 
sublattice 1.  This qualitative distinction between the $n = 0$ and 
$n \neq 0$ Landau
level wave functions has important consequences when we incorporate
interactions below.

We note that this feature of the $n = 0$ Landau level 
wave functions is robust against introducing spin-orbit
coupling, which admits a term ${\mathcal H}_{SO} = \Delta_{SO}\int
d^2{\bf x} \psi^\dagger \eta^z \sigma^z \psi$ to the Hamiltonian
\cite{SpinOrbit}.  In contrast, in the presence of this term the probability
density for $n \neq 0$ Landau level
wave functions will no longer be evenly distributed between both
sublattices.  This correction is quite small, as the
induced density difference between the sublattices will
be proportional to $\Delta_{SO}\ell_B/(\hbar v) \ll 1$, and will
hereafter be neglected.

\section{Integer quantum Hall ferromagnetism}
  
In this section we focus on a clean graphene system at integer filling
factors, and explore ``quantum Hall ferromagnetism'' using the
interacting theory derived above.  We will first analyze
the $n = 0$ Landau level, and then turn our attention to the $n = 1$
Landau level.  In the last part of this section we briefly discuss
current experimental relevance for our results.

\subsection{$n = 0$ Landau level}

We will employ two simplifying assumptions in our
analysis of quantum Hall ferromagnetism in the $n = 0$ Landau level.  
First, we make the standard approximation of ignoring
Landau level mixing and project out states away from the $n = 0$
Landau level.  This is reasonable given that the spacing
between the $n =0$ and $n = \pm 1$ Landau levels is roughly
$420\sqrt{B[{\rm T}]}$K, where $B[{\rm T}]$ is the magnetic field in
Teslas, while the Coulomb energy scale 
\begin{equation}
  {\mathcal E}_C \equiv \frac{e^2}{4\pi \epsilon_{\rm RPA} \ell_B} \approx 130 
  \sqrt{B[{\rm T}]}{\rm K}
  \label{Ec}
\end{equation}
 is around three times smaller.  Second, since the 
non-interacting wave functions live
on only one sublattice or the other, we will set
$\psi_{\alpha R/L 1} \rightarrow 0$.  
With the electron kinetic energy quenched, ${\mathcal H}_0$ then becomes
\begin{equation}
  {\mathcal H}_0^{n=0} = \frac{1}{2} \int d^2{\bf x}d^2{\bf x'} 
  \rho_{\rm tot}({\bf x}) V({\bf x-x'}) \rho_{\rm tot}({\bf x'}).
\end{equation}
The anisotropy terms in ${\mathcal H}_1$ are dramatically 
simplified upon projection,
\begin{eqnarray}
  {\mathcal H}_1^{n = 0} &=& \int d^2 {\bf x} \big{\{} -g \mu_B {\bf B}
  \cdot{\bf S}_{\rm tot}
  \nonumber \\
  &+&\frac{u_0}{2}\sum_{A = R/L}[\rho^2_{A2} -
  \frac{4}{3} {\bf S}^2_{A2}]
  \nonumber \\
  &-& \sum_{\bf r}v_1({\bf r})\rho_{\rm stag}({\bf
  x+r})\rho_{\rm stag}({\bf x})\big{\}}.
\end{eqnarray}
In the projected subspace, $\rho_{\rm tot} = \rho_{L2}+\rho_{R2}$, 
${\bf S}_{\rm tot} = {\bf S}_{L2}+{\bf S}_{R2}$, and 
$\rho_{\rm stag} = \rho_{L2}-\rho_{R2}$.  
Note that the $u_2$ interaction in Eq.\ (\ref{H1}) 
has dropped out altogether.  

In the SU(4)-invariant limit with ${\mathcal H}_1^{n=0} = 0$, despite the
four-fold degeneracy of the $n = 0$ Landau level the system is an exchange
ferromagnet at filling factors $\nu = -1$, 0, and $+1$ (corresponding
to a quarter-filled, half-filled, and three-quarter-filled level), and
exhibits a quantized Hall conductance $\sigma_{xy} = \nu e^2/h$.    
The SU(4) spin/flavor symmetry is broken spontaneously in the absence
of anisotropy terms, though in accordance with the Mermin-Wagner
theorem\cite{MerminWagner} 
true long-range order can exist only at zero temperature.
This spontaneous symmetry breaking gives rise to
three gapless Goldstone modes at $\nu = \pm 1$ and four at $\nu = 0$.
These SU(4) ``spin-waves'' have the following dispersion at small
momentum,\cite{Arovas}
\begin{equation}
  E_0({\bf q}) = \kappa (\ell_b q)^2,
\end{equation}
where the stiffness is given by
\begin{equation}
  \kappa = \frac{1}{4}\sqrt{\frac{\pi}{2}} {\mathcal E}_C.
\end{equation}
The lowest-energy 
charge excitations in these integer quantum Hall states 
are known to be SU(4) 
skyrmions\cite{Skyrmions, GirvinReview, SkyrmionsHigherLLs}, whose
energies are determined by the stiffness $\kappa$.  These topological
excitations were explored in Ref.\ \onlinecite{Arovas}.  Due to the Coulomb
repulsion the minimum-energy skyrmions are infinitely large, and the
energy cost for creating a skyrmion/antiskyrmion pair is 
\begin{equation}
  E_{sk/ask} = 2\kappa,
\end{equation} 
which is half that for an electron-hole pair.  

The goal of the remainder of this subsection will be to address how
this picture is modified by anisotropy terms in ${\mathcal H}_1^{n =
  0}$.  In particular, we will discuss the nature of the ground states
selected by these symmetry-breaking terms and examine their effects on
excitations out of the resulting ordered states.

\subsubsection{Filling factors $\nu = \pm 1$}

Consider now the $\nu = -1$ state, corresponding to a
quarter-filled $n = 0$ Landau level ($\nu = +1$ is related by
particle-hole symmetry and will not be discussed separately).  
Zeeman coupling favors a ground state occupied 
solely by spin up electrons.
Moreover, the $v_1$ term provides an ``easy-axis'' flavor symmetry that
favors having $\langle \rho_{\rm stag}\rangle \neq 0$ by occupying 
either all flavor $R$ or all flavor $L$ states.
It follows that in the presence of interactions the $\nu = - 1$ 
ground state exhibits a spontaneously broken $Z_2$ flavor symmetry,
and can therefore sustain long-range order at finite temperature.
Filling, say, the flavor $R$ states, the ground state
can be written
\begin{equation}
  |\nu = -1\rangle = \prod_{m} c^\dagger_{\uparrow R,m}|vac\rangle,
\end{equation}
where $|vac \rangle$ is the fermion vacuum and 
$c^\dagger_{\alpha A,m}$ adds an electron in the $n = 0$ Landau
level with spin $\alpha$, flavor $A$, and angular momentum $m$.
Since flavors $L$ and $R$ correspond to sublattices 1 and 2 in the
projected subspace, the $\nu = -1$ ground state is a spin-polarized
\emph{charge density wave} (CDW), having a larger electron density on
one of the two honeycomb sublattices.  The charge order is illustrated
schematically in Fig.\ \ref{honeycomb}.
Physically,  this CDW arises
because electrons can remain farther apart by occupying one sublattice
before the other, thereby minimizing their Coulomb energy.  
We emphasize,
however, that only the relatively small number of electrons
participating in
the $n = 0$ Landau level (roughly $1.4 \times 10^{-5} B[{\rm T}]$
electrons per hexagon) contribute to this order, which might make
it difficult to observe with an STM.  Another experimental signature
of the CDW at $\nu = -1$ may come from carbon NMR measurements.  
First, there is a net electron spin only on the higher-density sublattice, 
producing a magnetic moment per site of roughly 
$1.4 \times 10^{-5} B[{\rm T}] \mu_B$ with which the 
(relatively rare) carbon atoms with nuclear spin can interact. 
A second effect arises from more subtle lattice-scale structure that
exists in the $\nu = -1$ state.  Indeed, it
follows from Eqs.\ (\ref{DiracFields1}) and (\ref{DiracFields2}) 
that electrons on the higher-density
sublattice participate in currents circulating around second-neighbor
plaquettes, oriented along the arrows in Fig.\ \ref{honeycomb}.
This circulation in turn gives rise to a magnetic moment along the applied
field direction that couples to atoms in the \emph{lower-density}
sublattice.  
Using our zero-field relation between the lattice and continuum
fields, we crudely estimate the magnitude of this induced moment
to be $5 \times 10^{-5} B[{\rm T}] \mu_B$.
These effects should induce a small splitting in the nuclear spin
precession frequencies for carbon atoms on the two sublattices, which
may be measurable.

We turn next to excitations in the presence of anisotropy terms.  
The three Goldstone modes present in the
SU(4)-invariant limit become gapped.     
These generalized ``spin-wave'' gaps contain important information regarding
the stability of the spin and CDW order in the ground state, as well
as for the transport gap.
Employing the usual lowest Landau level projection, the energies for these 
three branches can be obtained essentially exactly.  We
follow closely the approach in the review in Ref.\
\onlinecite{GirvinReview}, generalized to the SU(4) case.  The
procedure is straightforward though tedious, and here we shall only
outline the calculation.  
Define 
\begin{equation}
  S^{\mu\nu} = \frac{1}{2} \psi^\dagger \tau^\mu \sigma^\nu \psi, 
  \label{SpinWaveOps}
\end{equation}
where $\mu, \nu$ run over 0, $x$, $y$, $z$ and $\tau^0$, $\sigma^0$ are
identity matrices.  
We wish to calculate the energies of
$\overline{S^{0x}}({\bf q})|\nu = -1\rangle$, $\overline{S^{x0}}({\bf
  q})|\nu = -1\rangle$, and $\overline{S^{xx}}({\bf q})|\nu =
-1\rangle$, which turn
out to be exact eigenstates of the projected interacting Hamiltonian.  
Here the overbar indicates a lowest Landau level 
projection, $\overline{S^{0x}}({\bf q})$
creates a spin-wave with momentum ${\bf q}$, 
$\overline{S^{x0}}({\bf q})$ creates a ``flavor-wave'', and 
$\overline{S^{xx}}({\bf q})$ creates a mixed spin/flavor-wave.  
The energies of these excited states are conveniently calculated in
first quantization.  The Hamiltonian in momentum space becomes
\begin{eqnarray}
  {\mathcal H}_0^{n=0} &=& \frac{1}{2} \int_{\bf k} V({\bf
  k})\overline\rho_{\rm tot}({\bf k})\overline\rho_{\rm tot}(-{\bf k})
  \\
  {\mathcal H}_1^{n=0} &=& -g \mu_B B \overline{S^{0z}}({\bf k} = 0)
  \nonumber \\
  &+& \frac{1}{4} u_0 \int_{\bf k}[\overline\rho_{\rm tot}({\bf
  k})\overline\rho_{\rm tot}(-{\bf k})
  + 4\overline{S^{z0}}({\bf k})\overline{S^{z0}}(-{\bf k})
  \nonumber \\
  &-&\frac{4}{3}(\overline{S^{zi}}({\bf k})\overline{S^{zi}}(-{\bf k})
  + \overline{S^{0i}}({\bf k})\overline{S^{0i}}(-{\bf k}))]
  \nonumber \\
  &-& 4\sum_{\bf r} v_1({\bf r})\int_{\bf k} e^{i{\bf k}\cdot
  {\bf r}} \overline{S^{z0}}({\bf k})\overline{S^{z0}}(-{\bf k}),
\end{eqnarray}
where $i$ is summed over $x, y, z$.  The projected first quantized
operators are given explicitly by
\begin{eqnarray}
   \overline\rho_{\rm tot}({\bf k}) &=& e^{-\frac{1}{4}(\ell_B k)^2} 
   \sum_j p_{\bf k}(j)
   \nonumber \\
   \overline{S^{\mu\nu}}({\bf k}) &=& \frac{1}{2} e^{-\frac{1}{4}
   (\ell_B k)^2}
   \sum_j p_{\bf k}(j) \tau^\mu(j) \sigma^\nu(j),
\end{eqnarray}
where $j$ runs over each electron in the ground state and $p_{\bf k}(j)$ 
is a unitary operator satisfying
\begin{eqnarray}
  p_{\bf k}(j) p_{\bf q}(j) &=& e^{i\frac{\ell_B^2}{2} k\wedge q} 
  p_{\bf k+q}(j)
  \\
  \ [p_{\bf k}(i), p_{\bf q}(j)] &=& 2i \delta_{ij}
  \sin(\ell_B^2 k \wedge q/2) p_{{\bf k+q}}(j)
\end{eqnarray}  
with $k \wedge q = {\bf \hat z}\cdot ({\bf k}\times {\bf q})$.
Evaluating the commutators $[{\mathcal H}_0^{n=0}+{\mathcal
    H}_1^{n=0},\overline{S^{0x}}({\bf q})]$, \emph{etc}.\ on the
ground state yields the corresponding excitation energies.  

At small momentum, we obtain the following energies for spin-wave,
flavor-wave, and mixed spin/flavor-wave excitations, respectively:
\begin{eqnarray}
  E_{\sigma}({\bf q}) &\approx& g\mu_B B + \kappa_\sigma (\ell_B q)^2
  \label{spinwave} \\ 
  E_{f}({\bf q}) &\approx& \sqrt{\frac{\pi^3}{24}}
  \bigg{(}\frac{a_0}{\ell_B}\bigg{)} u_1 \rho_0 + \kappa_f (\ell_B q)^2
  \label{flavorwave}
  \\
  E_{\sigma f}({\bf q}) &=& g\mu_B B + E_f({\bf q}),  
\end{eqnarray}
where $\rho_0 = 1/(2\pi \ell_B^2)$ and the stiffnesses are 
\begin{eqnarray}
  \kappa_\sigma &=& (u_0-u_1)\rho_0 + \kappa 
  \label{SpinStiffness}
  \\
  \kappa_f &=& u_1 \rho_0 + \kappa.
  \label{FlavorStiffness}
\end{eqnarray}
At large momentum $|{\bf q}|\rightarrow \infty$, these excitations
correspond to the three types of separated particle-hole excitations out of the
ferromagnetic state (particle-hole excitations can be created by
flipping a spin, flavor, or both).  Their energies are 
\begin{eqnarray}
  E_{\sigma}(\infty) &=& g\mu_B B + 2(u_0-u_1)\rho_0 + 4\kappa
  \label{spinP/Hexcitation} \\ 
  E_{f}(\infty) &=& 2 u_1\rho_0 + 4\kappa
  \label{flavorP/Hexcitation}
  \\
  E_{\sigma f}(\infty) &=& g\mu_B B +2u_1\rho_0 + 4\kappa.  
\end{eqnarray}

The most noteworthy feature of these excitation energies is the 
small value of the gap for flavor-wave excitations, which is $E_f(0) \approx
4\times 10^{-3} (B[{\rm T}])^{3/2}$K.
In fact, had we ignored the finite-range of $v_1$ here, we would
incorrectly conclude that the flavor-wave gap \emph{vanishes}.
Assuming a local form, the sublattice repulsion becomes
proportional to $\overline{S^{z0}}({\bf x})^2$, which is simply a
constant when acting on uniform density states in the quarter-filled
Landau level.  The lack of symmetry breaking from such local
interactions has been noted previously in studies of quantum Hall 
bilayers.\cite{Bilayer}
As a result the flavor wave gap is down by an additional factor of
$a_0/\ell_B$ from what one might naively anticipate.
Nevertheless, long-range CDW order is expected to
persist up to a transition temperature of order 
$\kappa_f/\ln[\kappa_f/E_f(0)]$.\cite{Dissipation}
The spin order is comparatively more robust, as the spin-wave gap
is $g\mu_B B \sim B[{\rm T}]$K,\cite{HighFieldExpt}
which for typical field ranges is around 50 times larger or so than the
flavor wave gap.  

The smallness of the flavor-wave gap has important implications
for skyrmions as well, which will continue to set the charge gap.  
In particular, skyrmions will be cheaper to create solely in 
flavor space, leaving the spin order intact.  The
competition between the sublattice repulsion $v_1$, which favors small
skyrmions, and the long-range Coulomb, which favors infinite
skyrmions, will set a length scale for the minimum-energy skyrmions.  
However, the ratio of the flavor-wave gap to the Coulomb energy scale
${\mathcal E}_C$ is quite small, given approximately by
$3\times 10^{-5} B[{\rm T}]$.  For comparison, the ratio of the Zeeman
gap to the Coulomb energy in GaAs is roughly $6 \times 10^{-3}
\sqrt{B[{\rm T}]}$, \cite{Fertig} which at 10T is around 60 times larger.  
Hence, corrections to the energies of flavor skyrmions arising from
the nonzero flavor-wave gap are expected to be small.  
Approximating their energies by those for infinite skyrmions in the
absence of a flavor-wave gap, the activation energy arising from
flavor skyrmions is estimated to be $\Delta_{\nu = -1}
\approx \kappa_f$.\cite{Skyrmions}
These results imply that the CDW order will diminish extremely rapidly
as one tunes away from $\nu = -1$, whereas the spin order
will be much more robust.  One possible test of these predictions would be
to measure the activation energy at constant perpendicular magnetic
field with varying in-plane fields \cite{ValleySkyrmions}.  The
activation energy arising from flavor skyrmions, being independent of spin,
will be insensitive to changes in the latter.

\subsubsection{Filling factor $\nu = 0$}

The ground state at $\nu = 0$, corresponding to a half-filled $n = 0$
Landau level, will depend on the strength of the Zeeman coupling and on-site 
$U$ relative to the sublattice repulsion $v_1$.  If the former terms
dominate, the ground state will be a spin-polarized flavor-singlet,
\begin{equation}
  |\nu = 0\rangle_{\rm sp} = \prod_m c^\dagger_{\uparrow
    R,m}c^\dagger_{\uparrow L,m}|vac\rangle,
\end{equation}
which in contrast to $\nu = -1$ respects all graphene lattice
symmetries.  If the sublattice repulsion dominates, 
however, the ground state will be a flavor-polarized spin-singlet, 
\begin{equation}
  |\nu = 0\rangle_{\rm fp} = \prod_m c^\dagger_{\uparrow
    R,m}c^\dagger_{\downarrow R,m}|vac\rangle,
\end{equation}
and thus also exhibit the lattice-scale CDW order illustrated in Fig.\
\ref{honeycomb}.  Comparing the energies of these two states, one
finds that the ground state will be spin-polarized if $g\mu_B B +
2u_0\rho_0>4 u_1\rho_0$, while a flavor-polarized ground state emerges
if $g\mu_B B + 2u_0\rho_0 < 4 u_1\rho_0$.  According to Eqs.\
(\ref{u0}) and (\ref{u1}), we have $u_0\rho_0 \approx 0.08 U[{\rm
    eV}]B[{\rm T}]$K and $u_1\rho_0 \approx 0.4 B[{\rm T}]$K,
where $U[{\rm eV}]$ is the strength of the on-site repulsion
evaluated in electron-volts.  Ascertaining which scenario
prevails is complicated by the rather large uncertainty in the value of
the on-site $U$.  Previous estimates have
suggested $U \sim$ 5-12eV,\cite{UestimateB} though here a slightly
lower range of values may be appropriate due to the presence of the
SiO$_2$ substrate.  With $g = 1.4$, which is the smallest value
reported in recent high-field experiments\cite{HighFieldExpt},
the critical value for $U$ above which the ground state is
spin-polarized is estimated to be $U_c \sim$ 4eV.  Hence we can 
not conclusively determine whether the ground
state is spin-polarized, though this regime can always be reached by
increasing the effective Zeeman coupling via the introduction of an
in-plane magnetic field component.   Below we will discuss excitations
in both regimes.  

Consider first the situation where the ground state is spin-polarized.  
Here there are four low-lying branches of excitations, generated by 
$\overline{S^{xx}}({\bf q}) \pm i\overline{S^{yx}}({\bf q})$ and 
$\overline{S^{0x}}({\bf q}) \pm \overline{S^{zx}}({\bf q})$, which
create spin waves by breaking flavor singlets as required by the Pauli
principle.  Schematically, the former breaks singlets by sending
$(|R\rangle|L\rangle -|L\rangle |R\rangle) \rightarrow |R\rangle
|R\rangle$ (or $|L\rangle |L\rangle$), while the latter sends 
$(|R\rangle|L\rangle -|L\rangle |R\rangle) \rightarrow |R\rangle
|L\rangle$ (or $|L\rangle |R\rangle$).  That is, 
$\overline{S^{xx}}({\bf q}) \pm i\overline{S^{yx}}({\bf q})$ creates a
spin-wave by doubly-occupying a given flavor, while  
$\overline{S^{0x}}({\bf q}) \pm \overline{S^{zx}}({\bf q})$ does not.
Proceeding as in $\nu = -1$, at small momentum the
respective energies for the first and second pairs of spin waves are 
\begin{eqnarray}
  E^{\rm sp}_{1\pm}({\bf q}) &\approx& g \mu_B B + 2u_0\rho_0-4u_1\rho_0 + \kappa_f
  (\ell_B q)^2
  \\
  E^{\rm sp}_{2\pm}({\bf q}) &\approx& g\mu_B B + \kappa_\sigma (\ell_B q)^2,
\end{eqnarray}
while at $|{\bf q}|\rightarrow \infty$ we have
\begin{equation}
  E^{\rm sp}_{1,2\pm}(\infty) = g\mu_B B + 2u_0\rho_0-2u_1\rho_0 + 4\kappa.
\end{equation}
The stiffnesses $\kappa_f$ and $\kappa_\sigma$ are the same 
as at $\nu = -1$.  Note that having $E^{\rm sp}_{1\pm}(0)>0$ requires that 
$g\mu_B B+ 2u_0\rho_0>4u_1\rho_0$, which is the same condition given
above for realizing a spin-polarized ground state.  

We will assume that $u_0-2u_1<0$, implying that 
$E^{\rm sp}_{1\pm}(0)<E^{\rm sp}_{2\pm}(0)$.  This occurs for $U\lesssim 10$eV,
which we believe is reasonable. 
Due to the smaller gap for exciting spin waves by doubly-occupying a
given flavor, it follows that the charge gap will be set by 
mixed spin/flavor-textured skyrmions.  Unfortunately, the large
uncertainty in the strength of the on-site $U$ does not permit us to
make a quantitative estimate for the activation energy set by these
skyrmions.  We can at least say qualitatively that as one tunes away 
from $\nu = 0$ they should induce a depletion in the total spin of the
system along with a simultaneous revival of the sublattice CDW order.
How rapidly this
occurs will of course depend on the optimal skyrmion size, which we
are unable to determine.  

Consider now the case where the sublattice repulsion dominates,
leading to a flavor-polarized ground state.  There are again
four low-lying branches of excitations, generated by 
$\overline{S^{xx}}({\bf q}) \pm i\overline{S^{xy}}({\bf q})$ and 
$\overline{S^{x0}}({\bf q}) \pm \overline{S^{xz}}({\bf q})$, which
create flavor waves by breaking spin singlets.  The first pair of
operators break singlets by sending
$(|\uparrow\rangle|\downarrow\rangle -|\downarrow\rangle
|\uparrow\rangle) \rightarrow |\uparrow\rangle
|\uparrow\rangle$ (or $|\downarrow\rangle |\downarrow\rangle$), 
while the latter operators send 
$(|\uparrow\rangle|\downarrow\rangle -|\downarrow\rangle
|\uparrow\rangle) \rightarrow |\uparrow\rangle
|\downarrow\rangle$ (or $|\downarrow\rangle |\uparrow\rangle$).  
The energies for the first and second pairs of flavor waves are
\begin{eqnarray}
  E^{\rm fp}_{1\pm}({\bf q}) &\approx& 
  \mp g \mu_B B + 2(2u_1-u_0)\rho_0 + \kappa_f (\ell_B q)^2
  \label{MinFlavorWaveGap}
  \\
  E^{\rm fp}_{2\pm}({\bf q}) &\approx& 2(2u_1-u_0)\rho_0 + 
  \kappa_f (\ell_B q)^2
\end{eqnarray}
at small momentum, while 
\begin{eqnarray}
  E^{\rm fp}_{1\pm}(\infty) &=& \mp g\mu_B B + 2(3u_1-u_0)\rho_0 + 4\kappa
  \\
  E^{\rm fp}_{2\pm}(\infty) &=&2 (3u_1-u_0)\rho_0 + 4\kappa.
\end{eqnarray}
Here the gap is
smallest for flavor waves exciting by doubly
occupying spin up states as favored by Zeeman coupling, so 
the charge gap will be set by mixed spin/flavor-textured
skyrmions in this case as well.  Away from $\nu = 0$ such skyrmions
will produce a rapid depletion of the CDW order and revival of 
spin polarization.
 
Distinguishing experimentally between these two ground states can be
achieved by measuring the activation energy in the presence of an in-plane
magnetic field that enhances the effective $g$-factor.  If the 
ground state is spin-polarized, then the
activation energy should increase with the in-plane field since the
larger $g$-factor favors smaller (and thus higher-energy) skyrmions.  On the
other hand, if the ground state is flavor-polarized, a
\emph{decrease} in the activation would be expected---at least
initially, before the condition for having a flavor-polarized ground
state is violated.  Here, the enhanced $g$-factor 
reduces the energy cost for slowly varying spin/flavor fluctuations 
according to Eq.\ (\ref{MinFlavorWaveGap}), leading
to an increase in optimal skyrmion size and hence a reduction in the
optimal skyrmion energy.

\subsection{n = 1 Landau level}

Next, we will discuss quantum Hall ferromagnetism at 
$\nu = 3,4$, and 5, corresponding to a quarter-filled,
half-filled, and three-quarter-filled $n = 1$ Landau level.  ($n = -1$
is related by particle-hole symmetry.)  We will again ignore Landau
level mixing and project onto the $n = 1$ level.  Since the wave
functions now reside on both sublattices, we must return to the full
interacting theory defined by Eqs.\ (\ref{H0}) and (\ref{H1}).  

In the SU(4)-invariant limit with ${\mathcal H}_1 = 0$, the system
will again be an exchange ferromagnet at filling factors $\nu = 3$, 4,
and 5.  It is interesting to explore the lowest-energy charge
excitations here, and in particular to ask whether the transport gap
is set by topological skyrmions or separated particle-hole
excitations.  In the case of Schrodinger equation Landau levels, it is
known that skyrmions constitute the minimum-energy charge excitation only in
the lowest Landau level.\cite{SkyrmionsHigherLLs}  Since we are
dealing here with two-component wave functions whose elements
consist of lowest Landau level and next-lowest Landau level
Schrodinger equation wave functions, the answer in the present case is
not \emph{a priori} obvious.  

To address this issue, we calculate the
energies of the generalized SU(4) ``spin-wave'' branches in the SU(4)-invariant
limit.  While the convenient first-quantized formalism described in
the previous subsection is no longer at our disposal, these energies
can be computed directly using second quantization.  For simplicity,
let us consider $\nu = 3$ and assume the following symmetry-broken
ground state,
\begin{equation}
  |0\rangle = \prod_m c^\dagger_{\uparrow R,m}|vac\rangle,
\end{equation}
where here $c^\dagger_{\uparrow R,m}$ adds an electron into the $n =
1$ Landau level.  (Our attention will be restricted to this filling factor
since the energies are unchanged at $\nu = 4$ and 5.)  
We evaluate explicitly the energy of spin wave
excitations created by the operator
$S^-_\sigma({\bf q}) \equiv \int_{\bf x} e^{-i {\bf q \cdot x}}[S^{0x}({\bf
    x}) - i S^{0y}({\bf x})]$, where $S^{0x}$ and $S^{0y}$ are defined
in Eq.\ (\ref{SpinWaveOps}).  This choice is arbitrary, since which branch we
evaluate is immaterial due to the assumed SU(4) symmetry.  
Upon projecting onto the $n=1$ Landau
level, we have $S^-_\sigma({\bf q}) \rightarrow 
\overline{S^-_\sigma}({\bf q})$, with
\begin{eqnarray}
  \overline{S^-_\sigma}({\bf q}) &=&
  \int_{\bf x} e^{-i {\bf q \cdot x}}\sum_{l,m}
  [\Phi^{n=1}_{l}({\bf x})]^\dagger\Phi^{n=1}_{m}({\bf x}) 
  \nonumber \\
  &\times& [c^\dagger_{\downarrow R,l}c_{\uparrow R,m} + 
  c^\dagger_{\downarrow L,l}c_{\uparrow L,m}].
\end{eqnarray}
Here $\Phi^{n = 1}_{l,m}({\bf x})$ are two-component wave functions
defined in Eq.\ (\ref{wavefunction}), with $l,m$ labeling the angular
momentum (we have suppressed the spin and flavor indices since the
wave functions do not depend on these).
A spin-wave in the projected space is thus given by $|{\bf q}\rangle 
= \overline{S^-_\sigma}({\bf q})|0\rangle$, and has an excitation energy
\begin{equation}
  E'({\bf q}) = \frac{\langle {\bf q}|\overline{{\mathcal H}_0}|{\bf
  q}\rangle}{\langle {\bf q}|{\bf q}\rangle} - 
  \frac{\langle 0|\overline{{\mathcal H}_0}|0\rangle}{\langle
  0|0\rangle},
\end{equation}
where $\overline{{\mathcal H}_0}$ is the SU(4)-invariant part of the
Hamiltonian projected onto the $n = 1$ Landau level.  Letting $L^2$ be
the system size, we obtain
\begin{eqnarray}
  E'({\bf q}) &=& 2\int_{\bf k}V(k) \sin^2(\ell_B^2 k\wedge q/2)N(k),
  \label{SpinWavesn=1}
  \\ 
  N(k) &=& \frac{2\pi}{L^2}\int_{\bf x y}e^{i{\bf k}\cdot({\bf
  x-y})}\sum_{l,m}[\Phi^{n=1}_l({\bf x})]^\dagger\Phi^{n=1}_m({\bf
  x})
  \nonumber \\
  &\times&[\Phi^{n=1}_m({\bf y})]^\dagger\Phi^{n=1}_l({\bf y})
  \nonumber \\
  &=& e^{-\frac{1}{2}(\ell_B k)^2}[1-(\ell_B k)^2/4]^2.
  \label{Normalization}
\end{eqnarray}
Note that replacing the wave functions in Eq.\ (\ref{Normalization})
by one-component Schrodinger equation Landau level wave functions 
reproduces the known spin-wave energies for the latter
case.\cite{SkyrmionsHigherLLs}  Thus, Eqs.\ (\ref{SpinWavesn=1}) and
Eq.\ (\ref{Normalization}) constitute a straightforward generalization
of the spin-wave energies to the case of two-component wave
functions. 

Equation (\ref{SpinWavesn=1}) yields the following energy dispersion
at small momentum,
\begin{eqnarray}
  E'_0({\bf q}) &\approx& \kappa' (\ell_B q)^2,
  \\
  \kappa' &=& \frac{7}{64}\sqrt{\frac{\pi}{2}}{\mathcal E}_C.
\end{eqnarray}
This implies that a skyrmion-antiskyrmion pair costs an energy
\begin{equation}
  E'_{sk/ask} = 2\kappa'.
\end{equation}
The energy of a particle-hole excitation meanwhile is given by
\begin{equation}
  E'_{p/h} = E'_0(\infty) =
  \frac{11}{16}\sqrt{\frac{\pi}{2}}{\mathcal E}_C.
\end{equation}
Comparing these, we see that $E'_{sk/ask}/E'_{p/h} = 7/22 < 1$;
consequently, in SU(4)-invariant limit skyrmions will set the transport gap
in the $n = 1$ Landau level as well.

\subsubsection{Filling factor $\nu = 3$}

We will now incorporate the anisotropy terms in ${\mathcal H}_1$ to 
deduce the probable ground states selected by interactions,
considering $\nu = 3$ first.  Due to the Zeeman coupling, we will assume
a spin-polarized ground state here.  The ordering in flavor space
remains to be determined.  The most general flavor-polarized state
favored by the long-range Coulomb repulsion can be expressed
\begin{equation}
  |\theta,\phi \rangle = \prod_{m =
   0}^\infty\bigg{[}\cos\frac{\theta}{2}
   c^\dagger_{\uparrow R,m} + \sin\frac{\theta}{2}
   e^{i\phi}c^\dagger_{\uparrow L,m}\bigg{]}|vac\rangle.
  \label{nu=3}
\end{equation}
The angles $\theta$ and $\phi$ specify the polarization 
direction in flavor space.  With the ground state spin-polarized, the
only terms in ${\mathcal H}_1$ remaining that break the 
flavor degeneracy are the $v_1$ and $u_2$ interactions.  Naively, the former
favors a state with $\langle \rho_{\rm stag} \rangle \neq 0$, while
the latter favors having $\langle J_\pm \rangle \neq 0$.  
It is easy to show, however, that both expectation values vanish for
\emph{any} choice of $\theta$ and $\phi$ in Eq.\ (\ref{nu=3}).  The
vanishing of $\langle \rho_{\rm stag} \rangle$ follows from the fact
that the $n = 1$ wave functions carry equal weight on both
sublattices.  The vanishing of $\langle J_\pm \rangle$ is a 
consequence of the fact that the elements
$\phi^{n=1}$ and $\chi^{n=1}$ 
in Eq.\ (\ref{wavefunction}) are given by
wave functions for different Schrodinger equation Landau levels; hence 
$\sum_m[\phi^{n=1}_m]^*\chi^{n=1}_m = 0$.  In other words, as opposed
to what we saw at $\nu = -1$, the \emph{direct} contributions from the
interactions leave the flavor symmetry unbroken here.  

There is, nevertheless, subtler flavor symmetry-breaking by
interactions that arises due to \emph{exchange}.  
Specifically, $v_1$ favors in-plane flavor polarization ($\theta =
\pi/2$), while $u_2$ favors out-of-plane flavor 
polarization (\emph{e.g.}, $\theta = 0$).  Denoting the
projected $v_1$ and $u_2$ interactions as 
$\overline{{\mathcal H}_{v_1}}$ and $\overline{{\mathcal H}_{u_2}}$, 
respectively, we obtain the following variational energies:
\begin{eqnarray}
  \langle \pi/2,\phi| \overline{{\mathcal H}_{v_1}}
  +\overline{{\mathcal H}_{u_2}} |\pi/2,\phi\rangle &=&
  \frac{\rho_0 L^2}{8} u_2\rho_0 
  \\
  \langle 0,\phi| \overline{{\mathcal H}_{v_1}}
  + \overline{{\mathcal H}_{u_2}}|0,\phi\rangle &=&
  \frac{\rho_0 L^2}{2} u_1\rho_0.
\end{eqnarray}
Since $u_1 \approx \sqrt{3} u_2/4$, it follows that the in-plane
polarized state has lower energy so that the $\nu = 3$ ground state is
given by
\begin{equation}
  |\nu = 3\rangle \sim \prod_{m =
   0}^\infty\big{[}c^\dagger_{\uparrow R,m} + 
   e^{i\phi}c^\dagger_{\uparrow L,m}\big{]}|vac\rangle,
\end{equation}
with a spontaneously chosen in-plane polarization angle $\phi$.
This state is characterized by a nonzero order parameter 
$\langle \psi^\dagger (\tau^x +  i \tau^y)\psi \rangle \neq 0$, and
in general breaks translation, rotation, and reflection lattice
symmetries for graphene.  Using Eqs.\ (\ref{DiracFields1}) and
(\ref{DiracFields2}), one can show that this discrete symmetry
breaking is due to lattice-scale fermion currents circulating around 
nearest-neighbor honeycomb plaquettes, with a specific pattern
determined by the polarization angle $\phi$.  

Since this state spontaneously breaks U(1) flavor symmetry, a
finite-temperature Kosterlitz-Thouless transition is expected here.
The physics associated with this transition has been examined in detail
previously in the context of quantum Hall bilayers, where the layer
degree of freedom plays the role of 
flavor.\cite{BilayerShort, Bilayer, BilayerReview}  
One obstacle for observing such a transition in the bilayer problem is
the finite tunneling amplitude between layers, which 
explicitly breaks the U(1) psuedospin symmetry that is analogous to 
the U(1) flavor symmetry in our case.  Here, however, the U(1) flavor
symmetry appears to be more robust.  Graphene may therefore 
eventually provide a clean
setting for studying the physics of a Kosterlitz-Thouless transition in
the quantum Hall effect.  Further theoretical exploration of this
phenomenon in the present context would certainly be interesting.

\subsubsection{Filling factor $\nu = 4$}

Similarly to what we found at $\nu = 0$, the ground state at $\nu = 4$
depends on the strength of Zeeman coupling and the on-site
$U$ relative to that of the interactions $v_1$ and $u_2$.  The former
terms again favor a spin-polarized flavor-singlet state, 
while the latter interactions now favor an in-plane flavor-polarized
spin-singlet.  Evaluating the variational energies of these states, 
we find that the spin-polarized state has a lower energy if 
$g\mu_B B + u_0 \rho_0 > u_1 \rho_0 + u_2\rho_0/4$.  This inequality 
is satisfied even with a vanishing on-site $U$ and a $g$-factor as
small as unity.  Thus we will assume here that $\nu = 4$ ground state
is spin-polarized:
\begin{equation}
  |\nu = 4\rangle = \prod_m c^\dagger_{\uparrow R,m}
   c^\dagger_{\uparrow L,m} |vac\rangle.
\end{equation}
We note, however, that the gaps for spin-waves out of this state are
positive only if $g\mu_B B+ 2u_0\rho_0>2u_1\rho_0 + u_2\rho_0/2$,
which provides a slightly more restrictive condition for having a
fully spin-polarized ground state.  With $g = 1.4$, the latter
inequality is satisfied provided $U \gtrsim 2$eV.

\subsubsection{Filling factor $\nu = 5$}

Finally, the situation at $\nu = 5$ is analogous to that at $\nu = 3$.
The ground state will have all spin up states in the $n = 1$ 
Landau level occupied to satisfy the Zeeman coupling.  Due to the 
sublattice repulsion $v_1$, the remaining spin down electrons
in the ground state are expected to be in-plane flavor-polarized:
\begin{equation}
  |\nu = 5\rangle \sim \prod_m \big{[}c^\dagger_{\downarrow R,m} + e^{i\phi}
   c^\dagger_{\downarrow L,m}\big{]} |\nu = 4\rangle.
\end{equation}
This state exhibits the same type of lattice-scale order as the $\nu =
3$ ground state, and also spontaneously breaks U(1) flavor symmetry so
that a Kosterlitz-Thouless transition is expected here as well.

\subsection{Experimental relevance?}

Initial sightings of the graphene quantum Hall effect observed integer 
quantum Hall states only at filling factors
$\nu = \pm 2, \pm 6, \pm 10$, \emph{etc}.,\cite{IQHE1,IQHE2} 
corresponding to filled
nearly four-fold degenerate Landau levels.  Quantum Hall ferromagnetism
was thus entirely absent in these experiments.  This is almost 
certainly due to a
collapse of the exchange gap responsible for quantum Hall
ferromagnetism by strong disorder, a phenomenon that has been established both
theoretically\cite{StonerCriterion, Sinova, GrapheneMacDonald} 
and experimentally\cite{ExchangeCollapse1, ExchangeCollapse2}.  

As discussed in the introduction, more recent experiments utilizing 
higher magnetic fields observed
additional quantized Hall plateaus at $\nu = 0, \pm 1$, and $\pm
4$.\cite{HighFieldExpt}  A natural question to ask is whether the
appearance of these plateaus is a manifestation of quantum Hall
ferromagnetism.  Nomura and MacDonald, who very recently derived an
experimental criterion for realizing quantum Hall ferromagnetism in graphene,
suggest that this is indeed the case.  However, the measured
activation energy at $\nu = \pm 4$ was found to be dominated by the
single-particle Zeeman splitting, suggesting that here
the system is perhaps more appropriately characterized as a 
\emph{paramagnet}.  At present it is unclear whether the
same is true for the plateaus at $\nu = 0$ or $\nu = \pm 1$.
Measurements of the activation energies there will likely provide some
hints as to their origin.  

In the following section, we will explore
the possibility that the system is indeed paramagnetic at these newly
observed quantized Hall plateaus, and that their origin is due to
explicit symmetry-breaking terms in the Hamiltonian (rather than
exchange) as appears to be the case at $\nu = \pm 4$.  In an attempt
to access this
regime, we will assume that the effects of exchange are effectively
``canceled'' by disorder, and utilize Hartree theory to examine
the spin- and flavor-splitting of the $n = 0$ and $n = 1$ Landau
levels by such terms.  This analysis will
hopefully provide useful input for determining the true nature of the
$\nu = 0$ and $\nu = \pm 1$ states.

\section{Paramagnetic Regime}

It will be instructive to begin our analysis of the paramagnetic
regime by applying Hartree theory initially in a clean system.  
Disorder effects will be discussed at the end of this section.  
Since in our treatment of the ferromagnetic regime we
found that the $u_2$ interaction dropped out at the $n = 0$ Landau
level and was effective only through exchange at $n = 1$, we will
ignore this term here.  Doing so will greatly streamline our discussion,
and our results do not depend on this simplification.  We retain the
remaining symmetry-breaking terms in ${\mathcal H}_1$, and consider
the following interacting Hamiltonian,
\begin{eqnarray}
  \tilde {\mathcal H} &=& \int d^2{\bf x}\big{\{}-i \hbar v 
  \psi^\dagger [\eta^x D_x+\eta^y D_y]\psi - g \mu_B {\bf
  B}\cdot {\bf S}_{\rm tot}
  \nonumber \\
  &+& u_0 [\rho_{\uparrow,{\rm tot}}\rho_{\downarrow,{\rm tot}}
  + \rho_{\uparrow,{\rm stag}}\rho_{\downarrow,{\rm stag}}]
  - u_1 \rho_{\rm stag}^2\big{\}},
  \label{Htilde}
\end{eqnarray}
where $\rho_{\alpha,{\rm tot}}$ and $\rho_{\alpha,{\rm stag}}$ are the
total and staggered densities for spin $\alpha$ (\emph{e.g.},
$\rho_{\uparrow,{\rm stag}} = \psi^\dagger_{\uparrow} \tau^z\eta^z \psi_{\uparrow}$).  We have ignored
the finite range of the sublattice repulsion $v_1$, as a local
form will now be adequate.  Furthermore, we have dispensed with the
manifestly SU(2)$_{\rm spin}$ invariant form of the on-site $U$ term in favor
of a form which will be more convenient to work with here.  Note also
that we have dropped pieces of the on-site $U$ interaction involving
$J_\pm$, since again these will be unimportant.  

Our aim will be to employ Hartree theory to obtain 
\emph{effective} single-particle energy levels
for the interacting system, and in particular to extract the 
resulting particle-hole
excitation energies which will be of potential relevance for
experimentally measured activation energies.  To proceed we decouple the 
interactions by linearizing in
fluctuations of the total and staggered densities about their mean values,
yielding an effective mean-field Hamiltonian
\begin{eqnarray}
  \tilde {\mathcal H}_{\rm MF} &=& \int d^2{\bf x}\big{\{}-i \hbar v 
  \psi^\dagger [\eta^x D_x+\eta^y D_y]\psi 
  \nonumber \\
  &-& g \mu_B {\bf B}\cdot {\bf S}_{\rm tot}
  + \sum_{\alpha = \uparrow,\downarrow}[\lambda_{\alpha,{\rm t}}
  \rho_{\alpha,{\rm tot}} + \lambda_{\alpha,{\rm s}}
  \rho_{\alpha,{\rm stag}}] 
  \nonumber \\
  &-&\mu_{\rm s} \rho_{\rm stag}\big{\}} + E_{\rm const}.
  \label{Hmf}
\end{eqnarray}
The constant-energy term that arises upon decoupling is given by
\begin{eqnarray}
  E_{\rm const} &=& \int d^2{\bf x}\big{\{}
  -u_0[\langle \rho_{\uparrow,{\rm tot}}\rangle
  \langle \rho_{\downarrow,{\rm tot}}\rangle 
  \nonumber \\
  &+& \langle \rho_{\uparrow,{\rm stag}}\rangle
  \langle \rho_{\downarrow,{\rm stag}}\rangle] 
  + u_1\langle \rho_{\rm stag}\rangle^2\big{\}}.
\end{eqnarray}
Furthermore, the new coupling constants appearing in Eq.\ (\ref{Hmf})
are defined as follows,
\begin{eqnarray}
  \lambda_{\uparrow/\downarrow,{\rm t}} &=& u_0 
  \langle \rho_{\downarrow/\uparrow,{\rm tot}}\rangle 
  \\
  \lambda_{\uparrow/\downarrow,{\rm s}} &=& u_0 
  \langle \rho_{\downarrow/\uparrow,{\rm stag}}\rangle 
  \\
  \mu_{\rm s} &=& 2 u_1 
  \langle \rho_{\rm stag}\rangle.
\end{eqnarray}
The $\lambda_{\alpha,{\rm t}}$ and $\lambda_{\alpha,{\rm s}}$ terms  
represent effective chemical potentials for
the total and staggered densities for spin $\alpha$, while $\mu_s$
behaves as an overall staggered chemical potential.
Assuming uniform expectation values for the total and staggered
densities, it is now straightforward to diagonalize the mean-field
Hamiltonian ${\mathcal H}_{\rm MF}$ to obtain the desired effective
single-particle levels for the interacting theory.  Below we examine 
the resulting level structure at 
integer filling factors in the $n = 0$ and $n = 1$ Landau levels.

\subsection{n = 0 Landau level}

The two-component $n = 0$ Landau level 
wave functions for flavor $A$ and spin $\alpha$ retain the same 
form as in the non-interacting case (see Sec.\
\ref{LLoverview}).  Their energies, however, are now shifted as
follows:
\begin{eqnarray}
  E_{\uparrow/\downarrow,R} &=& \mp \frac{1}{2}g\mu_B B +
  \lambda_{\uparrow/\downarrow,{\rm t}} -
  \lambda_{\uparrow/\downarrow,{\rm s}} + \mu_s
  \\
  E_{\uparrow/\downarrow,L} &=& \mp \frac{1}{2}g\mu_B B +
  \lambda_{\uparrow/\downarrow,{\rm t}} +
  \lambda_{\uparrow/\downarrow,{\rm s}} - \mu_s.
\end{eqnarray}
In particular, we see that the flavor degeneracy exhibited in 
the non-interacting case is generically lifted by interactions.
Below we specialize to filling factors $\nu = \pm 1$ and
$\nu = 0$ to obtain the effective level structure in the respective
ground states.  We will find here that Hartree theory predicts
the same ground states as in the ferromagnetic case, due to the fact
that direct contributions from interactions rather than exchange
selected the ordering in the latter case.  Excitation energies, however,
which are our main interest, will be dramatically modified by the 
loss of exchange.

\subsubsection{Filling factors $\nu = \pm 1$}

We will again consider only $\nu = -1$ since $\nu = +1$ is related by
particle-hole symmetry.  The mean-field energy at $\nu = -1$ is
minimized by filling all spin up, flavor $R$ states (or all spin up,
flavor $L$ states).  Setting $\langle \rho_{\uparrow,{\rm
  tot}} \rangle = -\langle \rho_{\uparrow,{\rm stag}}\rangle = \rho_0$ and 
$\langle \rho_{\downarrow,{\rm tot}} \rangle 
= \langle \rho_{\downarrow,{\rm stag}}\rangle = 0$, we obtain the following
effective single-particle levels:
\begin{eqnarray}
  E_{\uparrow R} &=& -\frac{1}{2} g\mu_B B - 2u_1 \rho_0
  \label{EupR}
  \\
  E_{\uparrow L} &=& -\frac{1}{2} g\mu_B B + 2u_1 \rho_0
  \\
  E_{\downarrow R} &=& +\frac{1}{2}g\mu_B B +2u_0\rho_0- 2u_1 \rho_0
  \\
  E_{\downarrow L} &=& +\frac{1}{2} g\mu_B B + 2u_1 \rho_0. 
  \label{EdownL}
\end{eqnarray}
By examining Eqs.\ (\ref{EupR}) through ({\ref{EdownL}), we see that the 
minimum energy required to make a
particle-hole excitation out of the ground state is either
$4u_1\rho_0$ or $g\mu_B B + 2 u_0\rho_0$, whichever is smaller.
Again we note that from Eqs.\ (\ref{u0}) and (\ref{u1}) we have
$u_0\rho_0 \approx 0.08 U[{\rm eV}]B[{\rm T}]$K and 
$u_1\rho_0 \approx 0.4 B[{\rm T}]$K, where $U[{\rm eV}]$ is expected
to be on the order of a few electron-volts.  
Figure \ref{EnergyLevels}(a) illustrates schematically the 
effective single-particle levels in these two cases.  

\begin{figure} 
  \begin{center} 
    {\resizebox{8cm}{!}{\includegraphics{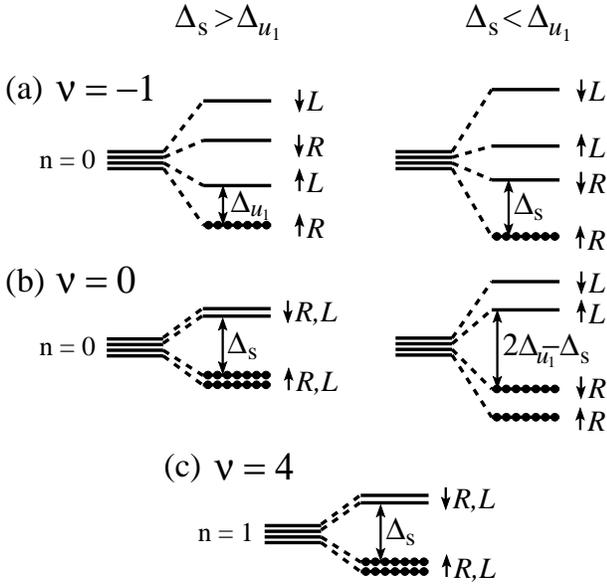}}} 
  \end{center} 
  \caption{Schematic \emph{effective} single-particle energy levels
    obtained within a Hartree analysis of the interacting theory at
    (a) $\nu = -1$, (b) $\nu = 0$, and (c) $\nu = 4$.
    Here $\Delta_s = g\mu_B B+ 2u_0 \rho_0$, while 
    $\Delta_{u_1} = 4 u_1\rho_0$.  For $\nu = -1$ and 0, the levels on
    the left-side correspond to $\Delta_s>\Delta_{u_1}$, while the
    right-side corresponds to $\Delta_s<\Delta_{u_1}$.} 
  \label{EnergyLevels} 
\end{figure}

\subsubsection{Filling factor $\nu = 0$}

Just as we saw in the ferromagnetic regime, the $\nu = 0$ ground state here
depends on the strength of Zeeman coupling and the on-site $U$
relative to the sublattice repulsion $u_1$.  Evaluating the mean-field
energetics for these states, we again find that the ground state will be
spin-polarized if $g\mu_B B + 2u_0 \rho_0>4u_1 \rho_0$, while a
flavor-polarized ground state exhibiting lattice-scale CDW order occurs 
if $g\mu_B B + 2u_0\rho_0< 4u_1 \rho_0$.  

Let us first deal with the case where the ground state is
spin-polarized.  Here, we set 
$\langle \rho_{\uparrow,{\rm tot}}\rangle = 2\rho_0$ and 
$\langle \rho_{\downarrow,{\rm tot}}\rangle =
\langle \rho_{\uparrow/\downarrow,{\rm stag}}\rangle = 0$, yielding
the following energy levels,
\begin{eqnarray}
  E^{\rm sp}_{\uparrow R/L} &=& -\frac{1}{2} g\mu_B B 
  \\
  E^{\rm sp}_{\downarrow R/L} &=& +\frac{1}{2} g\mu_B B + 2u_0 \rho_0.
\end{eqnarray}
Since all spin up states are occupied, the energy for a particle-hole
excitation is clearly given by $g\mu_B B + 2u_0\rho_0$.  
Consider alternatively the case where the ground state is
flavor-polarized, and we have 
$\langle \rho_{\uparrow/\downarrow,{\rm tot}}\rangle = 
-\langle \rho_{\uparrow/\downarrow,{\rm stag}}\rangle = \rho_0$
corresponding to occupying all flavor $R$ states.  The effective
single-particles levels are now
\begin{eqnarray}
  E^{\rm fp}_{\uparrow/\downarrow R} &=& \mp\frac{1}{2} g\mu_B B
  +2u_0\rho_0-4u_1\rho_0 
  \\
  E^{\rm fp}_{\uparrow/\downarrow L} &=& \mp\frac{1}{2} g\mu_B B + 4u_1 \rho_0.
\end{eqnarray}
Consequently, the minimum energy for a particle-hole excitation is
$8u_1\rho_0-2u_0\rho_0-g\mu_B B$.  
The single-particle levels for both scenarios are illustrated
schematically in Fig.\ \ref{EnergyLevels}(b).

\subsection{n = 1 Landau level}
  
We will make an additional simplification in our treatment of the $n =
1$ Landau level and set $\lambda_{\alpha,{\rm s}} = \mu_{\rm s} = 0$ 
at the outset.  This is not necessary and does not affect our results,
and we do so only for the sake of simplicity.
On physical grounds one expects these terms to vanish given that the
non-interacting $n = 1$ wave functions carry equal weight on both
sublattices.  Furthermore, it is easy to show that 
in any case they do not lead to a splitting of the
flavor degeneracy in the $n=1$ Landau level.  

With this simplification, the wave functions are unchanged from the
non-interacting case, while the energies for flavor $A$ become
\begin{equation}
  E'_{\uparrow/\downarrow A} = \mp \frac{1}{2} g\mu_B B +
  \lambda_{\uparrow/\downarrow,{\rm t}} + \sqrt{2e\hbar v^2 B}.
\end{equation}
In contrast to the $n = 0$ Landau level, the flavor degeneracy
remains unbroken by interactions here.  The important physical
implication is that there are gapless charge excitations at $\nu = 3$ and 5,
and thus no quantum Hall effect would be predicted by Hartree theory
at these filling factors.  This is not too surprising
since in our analysis of the ferromagnetic regime we found that
interactions broke the flavor degeneracy at $\nu = 3$ and 5 only via
exchange.  

The levels are spin-split, however, and there is a gap at $\nu = 4$.
All spin up orbitals are occupied in the ground state, so here we have
$\langle \rho_{\uparrow,{\rm tot}}\rangle = 2\rho_0$ and 
$\langle \rho_{\uparrow,{\rm tot}}\rangle =0$.  The effective
single-particle energies are thus given by
\begin{eqnarray}
  E'_{\uparrow A} &=& - \frac{1}{2} g\mu_B B + \sqrt{2e\hbar v^2 B}
  \\
  E'_{\downarrow A} &=& + \frac{1}{2} g\mu_B B + 
  2u_0\rho_0 + \sqrt{2e\hbar v^2 B},
\end{eqnarray}
and the particle-hole excitation energy is $g\mu_B B + 2u_0\rho_0$.  
The energy levels here are illustrated in Fig.\ \ref{EnergyLevels}(c).

\subsection{Discussion}

To summarize, our disorder-free Hartree analysis suggests that Zeeman
coupling together with symmetry-breaking interactions can \emph{in principle}
induce quantum Hall states in a paramagnetic system at filling 
factors $\nu = 0, \pm 1$, and $\pm 4$, but \emph{not} at 
$\nu = \pm 3$ or $\pm 5$.  The activation energies at $\nu = \pm 4$ 
obtained in Hartree theory---corresponding to one-half of the 
particle-hole energies---are 
\begin{equation}
  \Delta_{\nu = \pm 4} = \frac{1}{2}g\mu_B B + u_0\rho_0.
  \label{nu=4}
\end{equation}
At filling factors $\nu = \pm 1$ and $\nu = 0$, the activation
energies depend on whether ($i$) $g\mu_B B + 2 u_0 \rho_0 > 4 u_1
\rho_0$ or ($ii$) $g\mu_B B + 2 u_0 \rho_0 < 4 u_1 \rho_0$.  In case 
($i$) the activation energies at these filling factors are predicted to be
\begin{eqnarray}
  \Delta^{(i)}_{\nu = \pm 1} &=& 2 u_1 \rho_0,
  \label{nu=pm1i}
  \\
  \Delta^{(i)}_{\nu = 0} &=& \frac{1}{2}g\mu_B B + u_0\rho_0
  \label{nu=0i}
\end{eqnarray}
while for case ($ii$) we have
\begin{eqnarray}
  \Delta^{(ii)}_{\nu = \pm 1} &=& \frac{1}{2}g\mu_B B + u_0\rho_0.
  \label{nu=pm1ii}
  \\
  \Delta^{(ii)}_{\nu = 0} &=& 4u_1\rho_0-u_0\rho_0-\frac{1}{2} g\mu_B B
  \label{nu=0ii}
\end{eqnarray}
We note that in the second case the activation energy at 
$\nu = 0$ is larger than that for $\nu = \pm 1$ and $\pm 4$.  

These results will of course be modified by the inclusion of disorder.
In particular, one may ask whether the lattice-scale character of
the wave functions, which was essential for the flavor
symmetry-breaking at $n = 0$ and lack thereof at $n = 1$, survives
upon disordering the system.  We expect that microscopic features of the
wave functions will indeed remain robust, provided the length scale
associated with the disorder is long compared with the lattice
spacing.  Furthermore, it is clear that assuming uniform
expectation values for the total and staggered densities
as we did above will not be valid in the presence of
disorder.  This introduces a considerable degree of complexity, as the
mean-field Hamiltonian in this case is not easily diagonalized and
must be solved self-consistently.  
However, we expect that disorder can 
be effectively accounted for by a
(possibly magnetic field-dependent) broadening of the effective
single-particle levels found above, leading to a reduction in the
activation energies we predicted by considering a clean system.
A more serious treatment of disorder and interactions will clearly be
desirable, but will not be pursued here.

Even at the crude level of our treatment, it is interesting to compare
our predictions from Hartree theory with the high-field experiments
mentioned above that resolved additional quantum Hall states at 
$\nu = 0, \pm 1$, and $\pm 4$.  First, Hartree theory is consistent
with the appearance of these quantum Hall states, and also can potentially 
provide a simple explanation for the conspicuous absence of those 
at $\nu = \pm 3$ and $\pm 5$; namely, a lack of flavor 
symmetry-breaking due to the form of the higher-Landau level wave functions.  
Second, the activation energy predicted at $\nu = \pm 4$ is given by 
the Zeeman energy plus a contribution from the on-site $U$ that
is linear in the perpendicular magnetic field, the
latter being unchanged by the addition of an in-plane magnetic field
component [see Eq.\ (\ref{nu=4})].  The measured 
energy gap for the spin-split Landau levels was similarly found to be
given by the Zeeman energy, supplemented by a phenomenological Landau level 
broadening varying linearly with the perpendicular magnetic
field.\cite{HighFieldExpt}  If
one interprets this Landau level broadening as a contribution to the
energy splitting arising from the on-site $U$, one obtains $U \sim
2$eV, which is of the right order of magnitude.  Finally, 
if $g\mu_B B+2u_0\rho_0<4u_1\rho_0$, then one
obtains activation energies in Hartree theory which are identical at
$\nu = \pm 1$ and $\pm 4$, while a larger energy is predicted at $\nu =
0$ [see Eqs.\ (\ref{nu=4}), (\ref{nu=pm1ii}), and (\ref{nu=0ii})].  
It is intriguing
to note that experimentally the quantized
Hall plateau at $\nu = 0$ sets in at $B\approx 11$T, while the $\nu =
\pm 1$ and $\pm 4$ plateaus are resolved at similar fields around
$B\approx 17$T, which may be consistent with the relative magnitudes of 
the above Hartree energies. 

We emphasize that the preceding discussion 
is by no means intended to be conclusive, but only to suggest one
possible mechanism for the appearance of the additional quantum Hall
plateaus at high magnetic fields.  Again, assessing the origin of the 
quantum Hall states at $\nu = 0$ and $\nu = \pm 1$ requires further
experiments, which we hope this work may stimulate.  
Measuring the activation energies at these filling factors as a
function of an in-plane magnetic field may provide some guidance as to
whether the quantum Hall states at these filling factors are due to
quantum Hall ferromagnetism, explicit symmetry breaking, or perhaps
some other mechanism.  For instance, if 
$g\mu_B B + 2u_0\rho_0> 4u_1\rho_0$, our
Hartree estimates predict that the $\nu = 0$ gap will increase with an
in-plane magnetic field component, while the gaps at $\nu = \pm 1$
will remain unchanged.  On the other hand, for 
$g\mu_B B + 2u_0\rho_0<4u_1\rho_0$ the gaps at both $\nu = 0$ and $\nu
= \pm 1$ vary with an increasing in-plane field, the former 
\emph{decreasing} while the latter increases.

\section{Concluding Remarks}

As technological progress enables the fabrication of higher-quality
graphene samples, quantum Hall ferromagnetism will likely provide an
interesting avenue of exploration along the road to the fractional
quantum Hall effect.  One of the remarkable aspects of quantum Hall
ferromagnetism in graphene is the connection between quantum Hall
physics operating on long length scales of order the magnetic length
and lattice-scale physics.  The clearest manifestation of this
interplay occurs at filling factors $\nu = \pm 1$ and possibly also
$\nu = 0$, where lattice-scale charge density wave order, albeit weak, 
coexists with the integer quantum Hall effect.  Establishing the
presence of such lattice-scale structure experimentally would be
quite interesting,
although doing so will likely prove challenging as the signatures of 
such order are expected to be small.  Another remarkable aspect of
quantum Hall ferromagnetism in graphene that would be worth pursuing 
is the potential for observing a Kosterlitz-Thouless transition, which 
we argued should
occur at filling factors $\nu = \pm 3$ and $\pm 5$.  Further
experimental and theoretical studies of skyrmion physics in graphene 
would also be interesting.

In the more immediate future, further experiments to determine the
origin of the additional integer quantum Hall states appearing at 
high magnetic fields would be extremely useful.  We have provided
here Hartree estimates for the activation energies in these states
which, at least in principle, should be straightforward to compare
with by exploring the dependence of the transport gap on an in-plane
magnetic field.  Finally, studies that incorporate disorder into our 
analysis of the paramagnetic regime in a more systematic way would of
course be welcome.

\emph{Acknowledgments.} --- 
We would like to thank Andrei Bernevig, Taylor Hughes, and 
especially Leon Balents for
stimulating discussions, as well as Philip Kim for sharing 
experimental data prior to publication.  
This work was supported by the National Science Foundation
through grants PHY-9907949 (M.\ P.\ A.\ F.) and 
DMR-0210790 (J.\ A.\ and M.\ P.\ A.\ F.).


\end{document}